\documentclass[preprint,journal]{vgtc}            

\onlineid{1250}

\preprinttext{Accepted at IEEE Vis 2023. To appear in IEEE Transactions on Visualization and Computer Graphics.}

\vgtccategory{Research}

\vgtcpapertype{Evaluation}

\newcommand{\anj}[1]{{\color{black}#1}}

\title{Image or Information?  Examining the Nature and Impact\\ of Visualization Perceptual Classification}

\usepackage{url}
\usepackage{graphicx}
\usepackage{float}
\usepackage{amssymb}

\newcommand\sbullet[1][.5]{\mathbin{\vcenter{\hbox{\scalebox{#1}{$\bullet$}}}}}

\author{Anjana Arunkumar, Lace Padilla, Gi-Yeul Bae, and Chris Bryan}
\authorfooter{
$\sbullet$ Anjana Arunkumar, Gi-Yeul Bae, and Chris Bryan are with Arizona State University. Email: aarunku5@asu.edu, giyeulbae@asu.edu, cbryan16@asu.edu.\\
$\sbullet$ Lace Padilla is with Northeastern University, Email: l.padilla@northeastern.edu.
}

\abstract{
How do people internalize visualizations: as \textit{images} or \textit{information}? In this study, we investigate the nature of internalization for visualizations (i.e., how the mind encodes visualizations in memory) and how memory encoding affects its retrieval. \anj{This exploratory work examines the influence of various design elements on a user's perception of a chart}. Specifically, which design elements lead to perceptions of visualization as an \textit{image} (aims to provide visual references, evoke emotions, express creativity, and inspire philosophic thought) or as \textit{information} (aims to present complex data, information, or ideas concisely and promote analytical thinking)? Understanding how design elements contribute to viewers perceiving a visualization more as an image or information will help designers decide which elements to include to achieve their communication goals. For this study, we annotated 500 visualizations and analyzed the responses of 250 online participants, who rated the visualizations on a bilinear scale as `image' or `information.' We then conducted an in-person study (\textit{n} = 101) using a free recall task to examine how the image/information ratings and design elements impacted memory. 
The results revealed several interesting findings: Image-rated visualizations were perceived as more aesthetically `appealing,' `enjoyable,' and `pleasing.'  Information-rated visualizations were perceived as less `difficult to understand' and more aesthetically `likable' and `nice,' though participants expressed higher `positive' sentiment when viewing image-rated visualizations and felt less `guided to a conclusion.' The presence of axes and text annotations heavily influenced the likelihood of participants rating the visualization as `information.' We also found different patterns among participants that were older. Importantly, we show that visualizations internalized as `images' are less effective in conveying trends and messages, though they elicit a more positive emotional judgment, while `informative' visualizations exhibit annotation focused recall and elicit a more positive design judgment. We discuss the implications of this dissociation between aesthetic pleasure and perceived ease of use in visualization design.

}

\keywords{Information Visualization; Human-Centered Computing; Perception \& Cognition; Takeaways.}

\teaser{
  \centering
  \includegraphics[width=\linewidth]{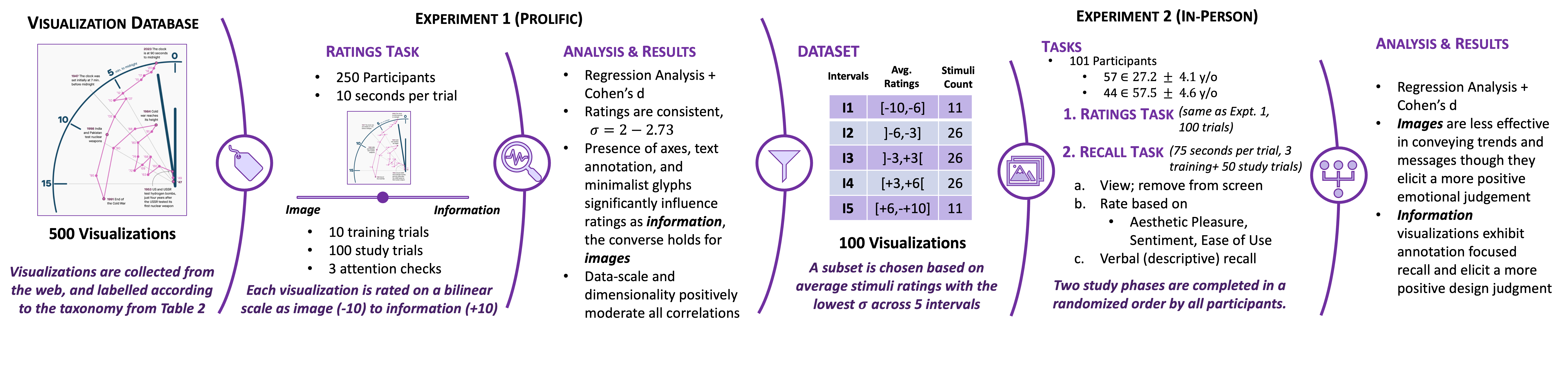}
  \caption{%
   Illustrative diagram of the experiments: 500 visualizations are labeled and categorized. In Experiment 1, \textit{image/information} ratings are gathered, and regression analysis and Cohen’s \textit{d} are used to determine the impact of different design elements of the charts on the ratings. In Experiment 2, a representative subset of 100 visualizations is curated,  and two tasks are performed: \textit{image/information} ratings and free recall; analysis compares the impact of ratings on the nature of the recall. 
  }
  \label{fig:teaser}
}

\graphicspath{{figs/}{figures/}{pictures/}{images/}{./}} 

\usepackage{tabu}                      
\usepackage{booktabs}                  
\usepackage{lipsum}                    
\usepackage{mwe}                       
\usepackage{soul}
\usepackage{mathptmx}                  

\begin{document}



\maketitle

\section{Introduction}

Communicative visualizations are rhetorical devices that inform and convince an audience of an idea by amplifying cognition~\cite{card1999using}. These represent the bulk of exposure any individual has to visualizations~\cite{9222102}. Scholars commonly define the \textit{purpose} of a communicative chart in terms of the \textit{intents} of a visualization designer~\cite{bertini2020shouldn}; these can be modeled as learning objectives, which can be \textit{cognitive} (analytical task performed) or \textit{affective} (emotional reaction elicited)~\cite{9905872} and dictate design choices.

Prior work has extensively examined the design elements that make some visualizations intrinsically more \textit{memorable} than others across charts in the wild~\cite{borkin2015beyond,borkin2013makes,franconeri2021science}. Concurrently, there have been attempts to define and evaluate higher-level concepts like \textit{engagement}~\cite{mahyar2015towards,boy2015storytelling,haroz2015isotype,moere2012evaluating} and \textit{effectiveness}~\cite{xiong2021visual,burlinson2020shape,arunkumar2021bayesian} across a plethora of (often synthetic and minimalist) chart types. However, the difference in the nature of charts studied through these different lenses poses challenges in reconciling good empirical results and rules of thumb with good exemplars of real-world visualizations. To disentangle these confounding factors, we set out to answer a fundamental question: \textit{how do people actually internalize visualizations: as image or information?}  We define \textit{images} as seeking to provide visual references, evoke 
emotions, express creativity, and inspire philosophic thought~\cite{10.2307/468718,john1997notebooks,KOSSLYN197752,10.2307/43029244,kosslyn1976can}. \textit{Information} visualization, on the other hand, focuses on presenting complex data, information, or ideas in a clear and concise manner, and promotes analytical thinking ~\cite{1320194,few2009now,kirk2012data,borkin2013makes,ware2019information,tufte1991envisioning} (discussion of definition in Section~\ref{sec:comVis}). While both use visual elements to convey meaning, they differ in their purpose and design.

In this paper, we present the results of two experiments (see the experimental design in Fig.\ref{fig:teaser}). We first annotated a dataset of 500 visualizations used ``in the wild" across social and scientific domains based on a taxonomy of static visualizations derived from literature~\cite{borkin2013makes,bryan2020analyzing} (see Table \ref{tab:1}). We then presented each visualization for 10 seconds to be rated on a bilinear scale as `image' or `information' (see Fig.\ref{fig:2} for definitions) by 250 online subjects. Based on response analysis, we took a representative sample of 100 visualizations from our database. We investigated the agreement between externalized ratings and internalized memory through a free recall task in a within-subjects experiment for 101 in-person subjects in two different age groups (20-35 y/o and 50+y/o). 
At a high level, the study results indicate consistency across age groups in rating patterns for the first experiment. For the second experiment, we observe similar descriptive patterns in the framing and focus of free recall verbiage for similar externalized rating values. Notably, we see that  the presence of axes and text annotations heavily influence the nature of internalization.

\textbf{Contributions:} This work represents the first study that systematically investigates which design elements of visualizations dictate the nature of internalization. To promote reproducibility, all study materials, visualization labeling and metadata, demographic details of study participants, study analysis and results, are publicly available at: \url{https://github.com/aarunku5/Image-or-Information-Vis-2023} (Note: anonymous repository is being used for review round). Based on the results of our experiment, we can offer quantitative evidence in direct support of several existing conventional qualitative visualization design guidelines promoting ease of use, including: (1) use of appropriate text annotation, (2) minimalist design style, and (3) presence of axes. We further identify a disconnect between aesthetic pleasure and perceived ease of use in visualization design, where \textit{image-like} visualizations may be less effective in conveying trends and messages, even though they tend to elicit a more positive emotional response. In contrast, \textit{informative} visualizations include more text, exhibit annotation-focused recall, and receive a more favorable design evaluation.

\section{Background}

\subsection{Perception and Memorability of Visualizations}

Broadly, the topic of graphical perception is concerned with how visualizations are perceived and interpreted~\cite{cleveland1987graphical}. Perception and readability depend on many factors, including the visual encodings being used (i.e., the marks and channels), the number of data dimensions being encoded~\cite{iliinsky2011designing,munzner2014visualization,padilla2020powerful}, the amount of data shown~\cite{keim2013big}, how the chart is styled \cite{kim2016data}, what rhetorical elements are present~\cite{hullman2011visualization}, and even the current cognitive focus of the person viewing the chart~\cite{healey2012attention}.

Previous work has focused on understanding the role of such visualization design choices on memorability and comprehensibility across a wide variety of charts~\cite{hullman2011benefitting,few2020chartjunk,tufte1985visual,isola2011understanding}. Additionally, several studies aiming to evaluate the impact of embellishments on visualization memorability and comprehension ~\cite{bateman2010useful,borgo2012empirical} have been linked to a ``slow analytics” movement that encourages ownership and retention of analytical tasks rather than precision~\cite{bradley2021approaching,lupi2017data}. Studies have demonstrated that some visualization types and visual elements are more memorable than others; for example, human-recognizable objects and color can increase the memorability of a visualization~\cite{borkin2013makes,borkin2015beyond}. 

In this study, we aim to move beyond memorability and investigate the nature of internalization for visualizations, i.e., how is a visualization encoded in memory (image/information rating), and how does the nature of memory encoding affect its retrieval (free recall verbiage patterns)? By identifying the most dominant features for the nature of rating and recall, designers can better understand how to reconcile empirical design guidelines with stylistic embellishments to improve the communicative efficacy of visualizations.

\subsection{Communicative Visualizations}
\label{sec:comVis}
Scholars often frame communicative visualizations in the context of ``learning about data''~\cite{fekete2008value,amar2005low,brehmer2013multi}, embedded in an analytical workflow of identifying and presenting an insight~\cite{viegas2007manyeyes,heer2007voyagers}. Research and practices focused on cognitive efficiency recommend the best visualization for accurate message decodings, such as generalized systems like APT and Tableau, which prioritize design based on specifications of the data of interest or broad analytical targets~\cite{kerpedjiev1998saying,mackinlay1986automating,pandey1912medley}.  However, in addition to visualizing data to support a claim accurately, design choices can appeal to emotion to persuade audiences to believe the data~\cite{buchanan1992wicked,conklin2005taxonomy,9222102}. For instance, researchers have utilized visualizations to bring attention to and provoke action for many issues~\cite{lambert2016and,roberts2015personal}. Designers can intentionally (or unintentionally) use conventions, such as minimalist aesthetics, the inclusion of data sources, or the increased use of text, to create an illusion of increased objectivity and transparency in such data representations for the audience~\cite{emerson2018challenging,kennedy2016work}. Notably, visual literacy plays a crucial role in determining the effectiveness of a communicative visualization in achieving such learning objectives for its target audience~\cite{9905872}.

\begin{figure}
    \centering
    \includegraphics[width=0.8\columnwidth]{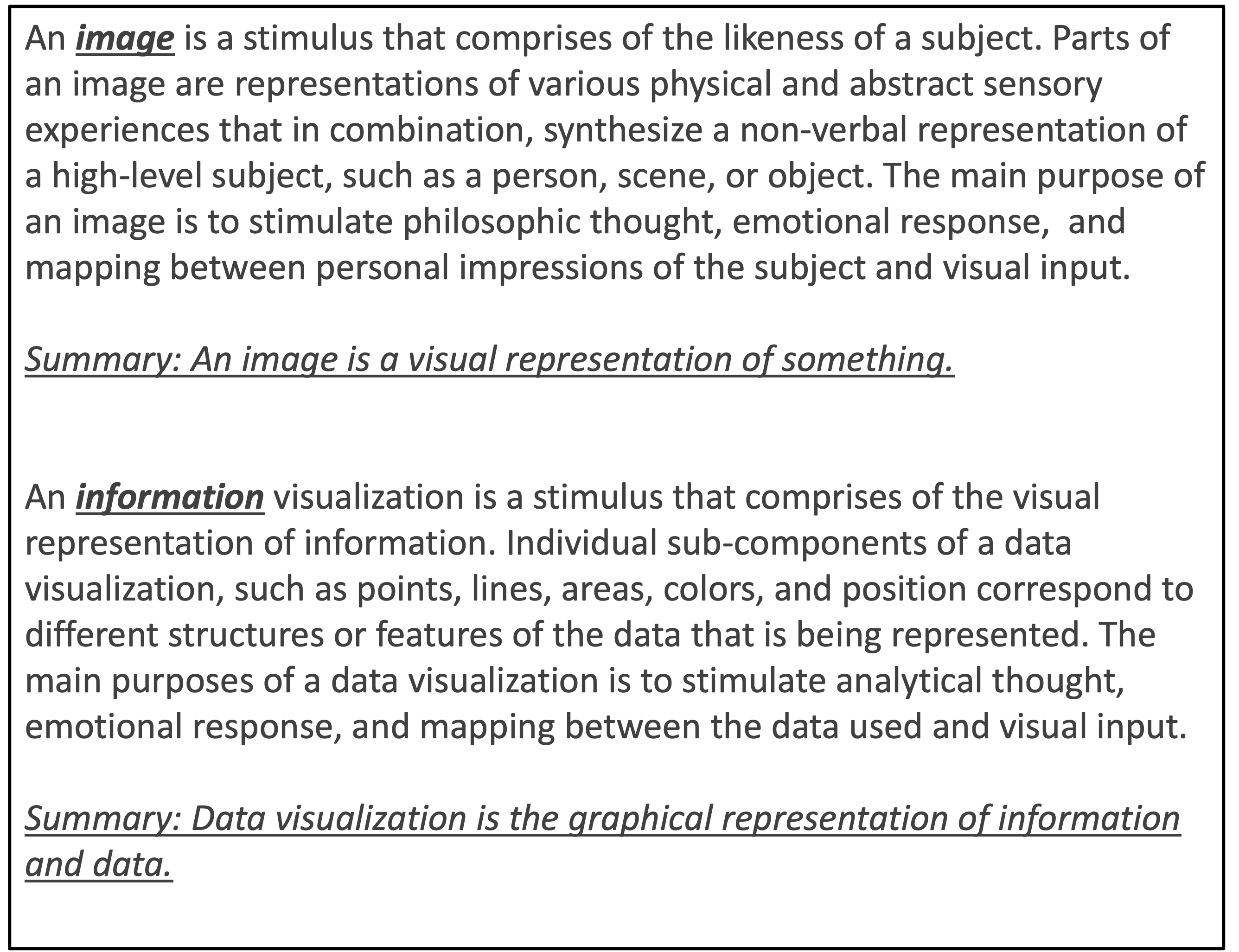}
    \caption{Constructed definitions of \textit{image} and \textit{information}.}
    \label{fig:2}
\end{figure}

 We construct definitions to clarify the nuance between `image' and `information' type visualizations (see Fig.~\ref{fig:2}) in the context of this communicative framework. Although both use visual components to communicate a message, their purpose and style vary. Our definitions are curated from a synthesis of literature~\cite{10.2307/468718,john1997notebooks,KOSSLYN197752,10.2307/43029244,kosslyn1976can,1320194,few2009now,kirk2012data,borkin2013makes,ware2019information,tufte1991envisioning,arnheim1969art}, and consider: (i) \textit{design}: images often prioritize aesthetics, composition, and visual impact, while information is typically presented in a structured, organized manner to facilitate understanding and retention, (ii) \textit{purpose}: images visually communicate an idea or evoke emotional engagement, while information educates, informs, or instructs the viewer, and (iii) \textit{response}: images seek to elicit visceral, emotional responses and rely on intuitive processing, while information facilitates the viewer's analysis/understanding of complex topics, and promotes long-term recall.

\subsection{Infographics and Relationships with Data Visualization}

Infographics are a form of data representation primarily intended for effective communication. In comparison, data visualizations can be used not just to communicate but also for exploration, discovery, and analysis, among other tasks~\cite{burns2021designing,cairo2016truthful}. Outside the visualization community, infographics are used to promote audience engagement~\cite{houts2006role} and comprehension of information~\cite{dick2014interactive}, particularly  for individuals with low-to-medium graph literacy~\cite{burns2021designing,garcia2013communicating}.

Past research has demonstrated increased memorability of infographics compared to unembellished charts~\cite{bateman2010useful,borkin2013makes,borkin2015beyond}, with a more accurate short-term recall of chart message/data points~\cite{franconeri2021science,harrison2015infographic} for colorful, non-complex representations. Some studies find infographics are less effective than other narrative visualization techniques like data comics~\cite{ovans2014makes,wang2019comparing} and primarily serve as a curated exploration method rather than an analytical tool. Additionally, studies have shown that the internal memory of a visualization is influenced by a viewer's impression of a canonical representation of that visualization type~\cite{franconeri2021science,tversky1989perceptual}.

The visualizations used in our study are within the purview of static infographics (with varying levels of text annotation). We analyze which design elements in such infographics detract the most from perceived ease of use, i.e., when the visualization is rated as highly image-like, for study stimuli and their relationship with aesthetic appreciation.

\begin{table*}[!ht]
\centering
\scriptsize
\resizebox{0.9\textwidth}{!}{%
\begin{tabular}
{p{0.1\textwidth}p{0.42\textwidth}p{0.38\textwidth}p{0.1\textwidth}}
\toprule
\textbf{Source} & \textbf{Websites} & \textbf{Topics} & \textbf{Total (500/\underline{100})}\\\midrule
Government     & Department of Energy, Los Angeles Planning Commission, Nasa, The Struggle for Five Years in Four , Washington Census, White House Office of Management and Budget, US Treasury Dept., World Health Organization (WHO)                              & Demographics, Economics, Environment, Health, Power, Traffic, Space       & 28/\underline{5}                                  \\
News                          & BBC, Bloomberg, Chicago Tribune, Economist, Guardian, Huff Post, Microsoft, NBC, National Post, New York Times, Quartz, Sports Reference, Visual Capitalist, Wall Street Journal, Washington Post & Demographics, Economics, Food, Geography, Geology, Health, Language, Politics, Pop Culture, Power, Resources, Space, Sports, Technology, Traffic, Weather       & 112/\underline{22}                                  \\
Media                       & 3CS.CH, Beautiful Evidence, Behance, Bike MS, DataViz Catalogue, Envisioning Information, Exploring Data, Facebook, FiveThirtyEight, Flickr, Forbes, Fortune, Fundamentals of Data Visualization, GoIreland, Hive Systems, Mapbox, NFL, NPM-JS, Neon, Observable, Reuters, Selfiecity, Tech in Review, The Pudding, Velir, Venngage, Verilogue, Visual Capitalist                                                                     & Demographics, Economics, Environment, Food, Geography, Geology, Health, History, Language, Lifestyle, Miscellaneous, Politics, Pop Culture, Power, Resources, Science, Sports, Technology, Traffic, Weather       & 52/\underline{15}                                  \\
Blogs                  & Adventures in Mapping, Flowing Data, Giorgia Lupi, Matt Mckeon, Midwest Uncertainty Collective, Multiple Views, Nicolas Rapp, Nightingale, R-Graph Gallery, Towards Data Science, Vis Guides, Visual Cinnamon                              & Demographics, Economics, Environment, Food, Geography, Geology, History, Health, Language, Lifestyle, Military, Politics, Pop Culture, Power, Resources, Science, Space, Sports, Technology, Traffic, Weather                    & 86/\underline{21}                                  \\
Social Media                 & Flickr, Pinterest, Reddit, Twitter                              & Demographics, Economics, Environment, Food, Geography, Geology, Health, History, Language, Lifestyle, Miscellaneous, Politics, Pop Culture, Power, Resources, Sports, Technology, Traffic, Weather            & 56/\underline{13}                                  \\
Infographics                        & Beautiful News, Visual.ly, Information is Beautiful, Tableau Public                                                                     & Culture, Demographics, Economics, Environment, Food, Geography, Geology, Health, History, Language, Lifestyle, Miscellaneous, Politics, Pop Culture, Power, Resources, Science, Sports, Technology, Traffic, Weather       & 137/\underline{22}                                  \\
Scientific Publications             & IEEE, Industrial Ecology, Nature, OpenCelliD, Oxford, Perspecta, R-chie, University of Santa Clara, Vassar                                                                        & Demographics, Economics, Geography, Geology, History, Military, Resources, Science, Technology, Traffic, Weather                & 29/\underline{2}           \\ \bottomrule                          
\end{tabular}%
}
\caption{List of visualization sources, their topics, and the respective number of visualizations from the full corpus (500) / \underline{subset (100)}.}
\label{tab:2a}
\end{table*}

\subsection{Visualization Aesthetics}

Prior work across domains like advertising and market design has demonstrated that high ratings of aesthetic quality are associated with greater visual exploration and perceived usability~\cite{maughan2007like,tractinsky2000beautiful}. However, these works use stimuli that cannot be categorized as infographics or visualizations, though they may comprise part of such elements. In the visualization context, \textit{aesthetics} refers to a quality or characteristic of a visual representation distinct from how clear, informative, or memorable it is. An alternate definition refers to the visual appeal or beauty of the representation~\cite{he2022beauvis,di2023doom}. We seek to systematically evaluate the applicability of the premise that beauty and functionality are intrinsically intertwined~\cite{bateman2010useful,borkin2013makes,chen2005top,healey2004perceptually,healey2012attention,hogarth1753analysis} for visualization. Is what is useful often beautiful? Is what is beautiful often useful? In our free recall task, we ask participants to explicitly rate aspects of aesthetic pleasure based on emotional response and design quality considerations using the Beauvis scale~\cite{he2022beauvis} (see Fig. \ref{fig:3b}). We compare these ratings with the image/information scores for stimuli, as well as ratings on perceived ease of use (i.e., participants' `experiencing difficulty in understanding' and `feeling guided to a conclusion' by stimuli).

\subsection{Visualization Engagement}

Visualization researchers have offered varying definitions of \textit{user engagement}. Some define user engagement as the willingness of users to invest effort in exploring a visualization to gain more information~\cite{haroz2015isotype,boy2015storytelling}, while others define it as the perceived effectiveness of visualizations in terms of participation or use~\cite{moere2012evaluating}. Different aesthetic styles of visualization, such as `sketchy'~\cite{wood2012sketchy}, `magazine'~\cite{moere2012evaluating}, `pictographic'~\cite{haroz2015isotype}, and `analytical'~\cite{moere2012evaluating}, and interaction measures such as hover~\cite{boy2015storytelling}, clicks~\cite{haroz2015isotype}, and willingness to annotate~\cite{wood2012sketchy} have been analyzed for user engagement. However, factors such as users’ goals, interest, and familiarity of the data, as well as accessibility and display parameters, can affect their willingness to explore the visualizations and also eventually determine the insights they gain~\cite{yi2008understanding}.

Mahyar et al.~\cite{mahyar2015towards} propose a five-level taxonomy for engagement (Expose--Involve--Analyze--Synthesize--Decide) based on prior literature~\cite{bloom2020taxonomy} which posits that the degree of
engagement increases as a user performs higher-level cognitive tasks such as synthesizing information, and making
final decisions.  To further ground this work in current visualization efforts, we measure the agreement between participants' internalization ratings (i.e., \textit{image/information?}) and the elements of visualizations they focus on during free recall (i.e., what is mentioned and what is emphasized) in the context of this taxonomy.


\begin{figure}[h]
\centering
\includegraphics[width=0.8\columnwidth]{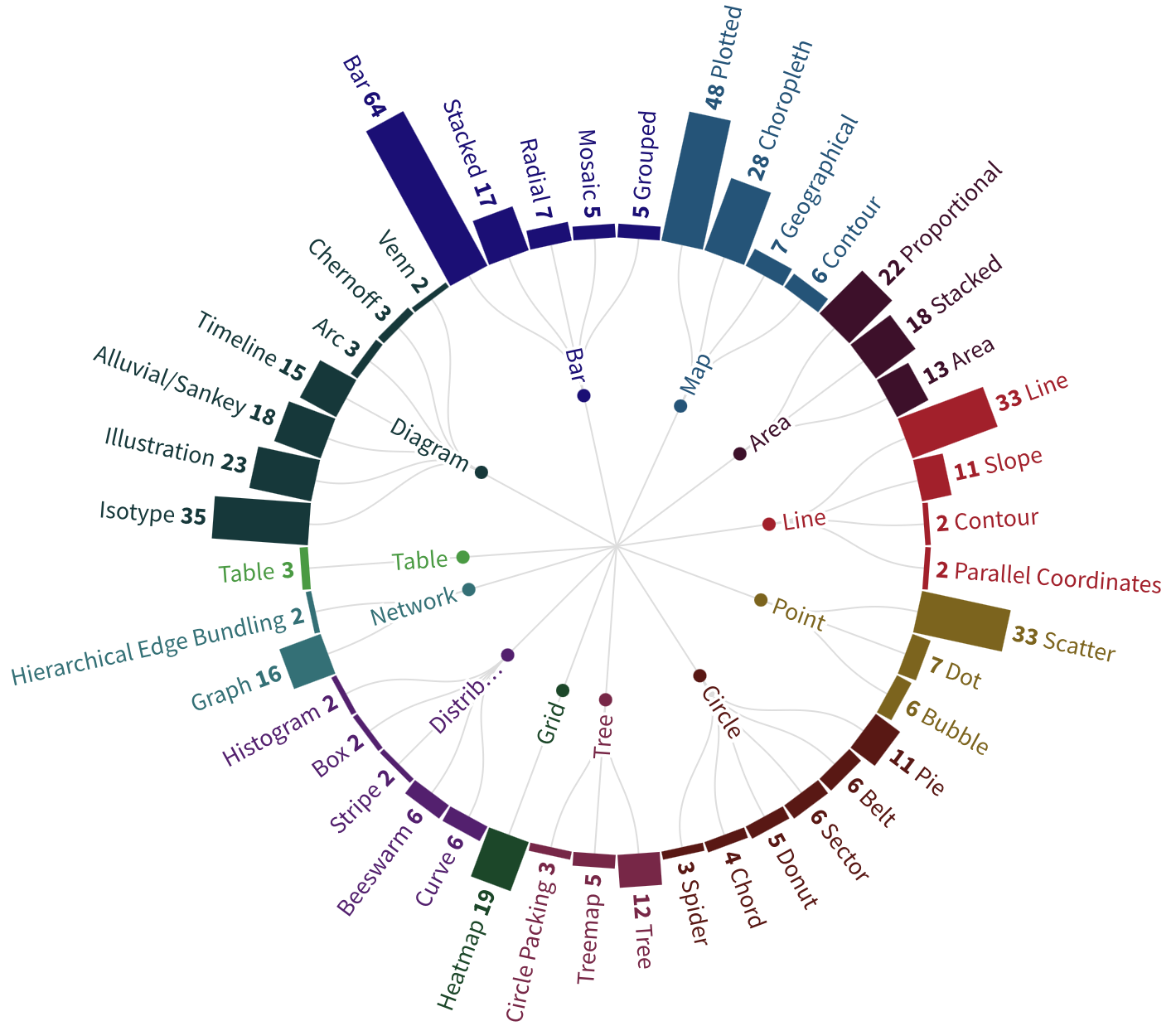}
\caption{Count for different visualization types in the database.}
\label{tab:2b}
\end{figure}

\section{Stimuli Overview} \label{sec:3}

To obtain many real-world examples encompassing a breadth of visualization types, design aesthetics, and domains, we initially collected 500 data visualizations (for \hyperref[sec:expt1]{Experiment 1}) from multiple sources, as shown in Table \ref{tab:2a} and Fig.~\ref{tab:2b}. These particular sources were chosen in line with data collection efforts like MASSVIS.\footnote{\url{http://massvis.mit.edu/}.} In total, we identified 460 \textit{single} visualizations, i.e., stand-alone visualizations with one panel, and 40 \textit{multi-panel} visualizations, that contain multiple related visualizations as part of a single narrative. We note that some \textit{single} visualizations may use other visualizations as glyphs (for instance, maps can be overlaid with pies/bars/networks across various geographic regions of interest).

\begin{table*}
    \centering
\resizebox{\textwidth}{!}{%
\begin{tabular}{lll}
\toprule
\textbf{Properties}                       & \textbf{}                                                                                                                                                                           & \textbf{Possible Values}                                                          \\ \midrule
\textbf{Dimension}                        & the number of variables being represented                                                                                                                                           & \textit{low (\textless{}3) / medium (3-4) / high (\textgreater{}4)}                                                               \\
\textbf{Multiplicity}                     & is the visualization stand-alone or somehow grouped with other visualizations                                                                                                       & \textit{single/multi-panel}                                                       \\
\textbf{Warping}                          & is the visualization radial in nature or have curvilinearly styled axes                                                                                                             & \textit{y/n}                                                                      \\
\textbf{Temporal}                         & is the visualization representing time-series data                                                                                                                                  & \textit{y/n}                                                                      \\
\textbf{Pictorial}                        & encoding is a pictogram; individual pictograms represent units of data such as in Isotype visualizations                                                                            & \textit{y/n}                                                                      \\ \midrule
\multicolumn{2}{l}{\textbf{Attributes}}                                                                                                                                                                                         &       \textbf{Possible Values}                                                                            \\ \midrule
\textbf{Black and White}                  & is the visualization only in gray-scale                                                                                                                                             & \textit{y/n}                                                                      \\
\textbf{Number of Distinct Colors}        & how many distinct color hues are present                                                                                                                                            & \textit{2-7}                                                                      \\
\textbf{Data Ink Ratio}                   & the ratio of data to non-data elements                                                                                                                                              & \textit{low/medium/high}                                                          \\
\textbf{Visual Density}                   & overall density of visual elements in the image without distinguishing between data and non-data elements                                                                           & low/medium/high                                                                   \\
\textbf{Skeuomorphism}                    & includes some element or has an overall appearance which resembles a real-world object or aligns with the & \textit{y/n}\\
& broad topic of the chart, though its rendering may not be photo-realistic &                                                                     \\
\textbf{Human Recognizable Objects (HRO)} & includes skeuomorphic glyphs/annotations, logos, symbols                                                                                                                            & \textit{y/n}                                                                      \\
\textbf{Human Depiction (HD)}             & presence of a photo-realistic or skeuomorphic representation of a human                                                                                                             & \textit{y/n}                                                                      \\
\textbf{Glyph/Overlay}                    & presence of a glyph/overlay element in the visualization                                                                                                                            & \textit{circles/squares/planes/food/etc.}                                         \\
\textbf{Ordering}                         & how are visualization elements ordered, if applicable                                                                                                                               & \textit{increasing/decreasing/temporal/random/nan}                                \\
\textbf{3D}                               & is the visualization depicted as a 3-dimensional figure                                                                                                                             & \textit{y/n}                                                                      \\
\textbf{Data Volume}                      & what is the scale of data being visualized                                                                                                                                          & \textit{low (\textless{}10) / medium (\textless{}100) / high (\textgreater{}100)} \\
\textbf{Background Color}                 & what kind of background color hue is present                                                                                                                                        & \textit{white/dark/light}                                                         \\
\textbf{Text Volume}                      & what is the volume of text present in the visualization (based on word count)                                                                                                       & low (\textless{}10) / medium (\textless{}25) / high (\textgreater{}25)            \\
\textbf{Data Source}                      & is the data source specified within the visualization                                                                                                                               & \textit{y/n}                                                                      \\
\textbf{Photo-Realism}                    & includes some element which depicts a real-world object with the exactness of a photograph                                                                                          & \textit{y/n}                                                                      \\
\textbf{Sentiment}                        & what is the emotional response evoked on viewing                                                                                                                                    & \textit{positive/negative/neutral}                                                \\
\textbf{Purpose}                           & what is the level of engagement expected from a viewer                                                                                                                              & \textit{expose/involve/analyze}                                                   \\ \midrule
\multicolumn{3}{l}{\textbf{Presence/Absence:} Axes, Title, Key, Caption, Aggregate Data, Data Redundancy, Message Redundancy, Grid Lines, Annotations of Interest (AOI) -  Text, Lines, Arrows, Highlights}                         \\ \bottomrule                                                             
\end{tabular}%
}
\caption{Summary of classified chart elements, properties, attributes.}
    \label{tab:1}
\end{table*}

A higher proportion of the visualizations collected are from infographic sources or social media. These sites are more likely to be accessed and encountered by non-expert users; such sources have more `diagrams.' Skeuomorphism is also more commonly seen in these sources, where the chart's overall structure or individual elements mimic real-world counterparts relevant to its topic/data domain. `Maps' are highly represented across all sources except for scientific publications, which, along with government sources, use basic visual encoding techniques, such as line graphs, bar charts, and point plots. Overall, infographic and scientific sources have a relatively larger proportion of \textit{multi-panel} charts compared to other sources. 

We manually labeled visualizations using 5 `expert' annotators (graduate students with 3+ years of research experience in data visualization) based on the visual taxonomy developed by Borkin et al.~\cite{borkin2013makes}, which classifies visualizations based on underlying data structures, visual encodings used, and the perceptual tasks enabled by these encodings. \anj{All coders annotated the full corpus, and majority coder agreement was used for cases of disagreement} 
Metadata and labeling information for the dataset can be found in the anonymized repository that hosts supplemental material. Additionally, our experts annotated the stimuli based on a set of constraints that may apply to any of the visualization types (see Table \ref{tab:1}), drawing on prior work~\cite{borkin2013makes, bryan2020analyzing}.

\section{Experiment Overview} \label{{sec:5}}
\subsection{Set-up and Participants} 

\begin{figure}[H]
  \centering
  \subfloat[Ratings Task]{\includegraphics[width=0.35\textwidth]{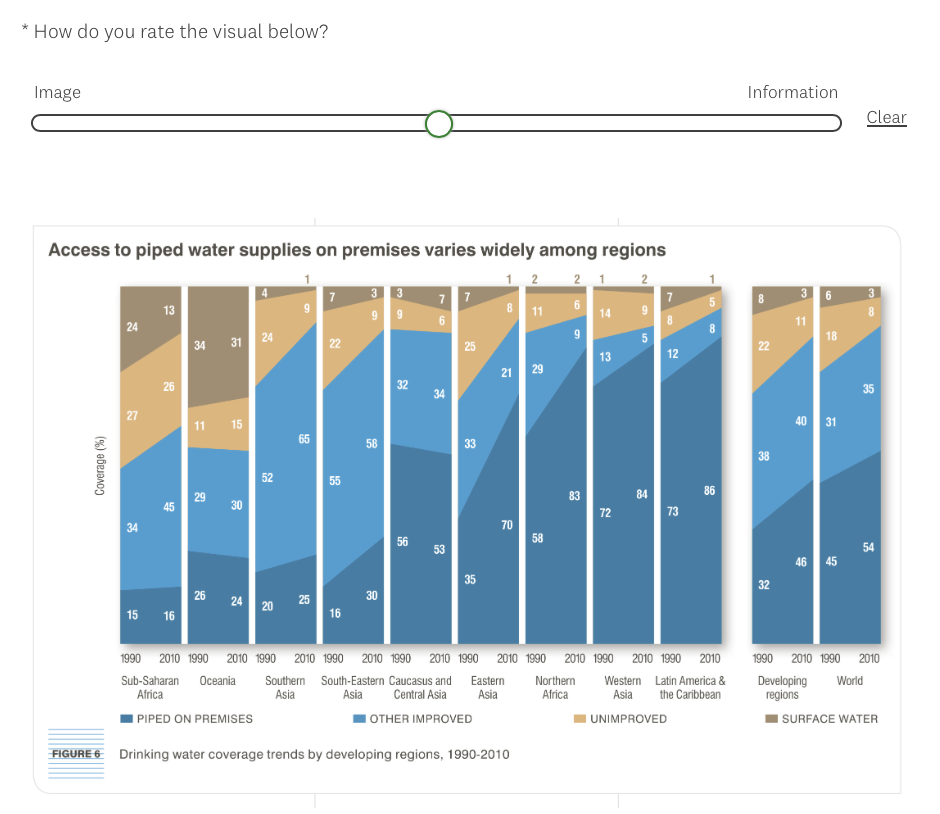}\label{fig:3a}}
  \hfill
  \subfloat[Recall Task (+ Verbal Recall)]{\includegraphics[width=0.35\textwidth]{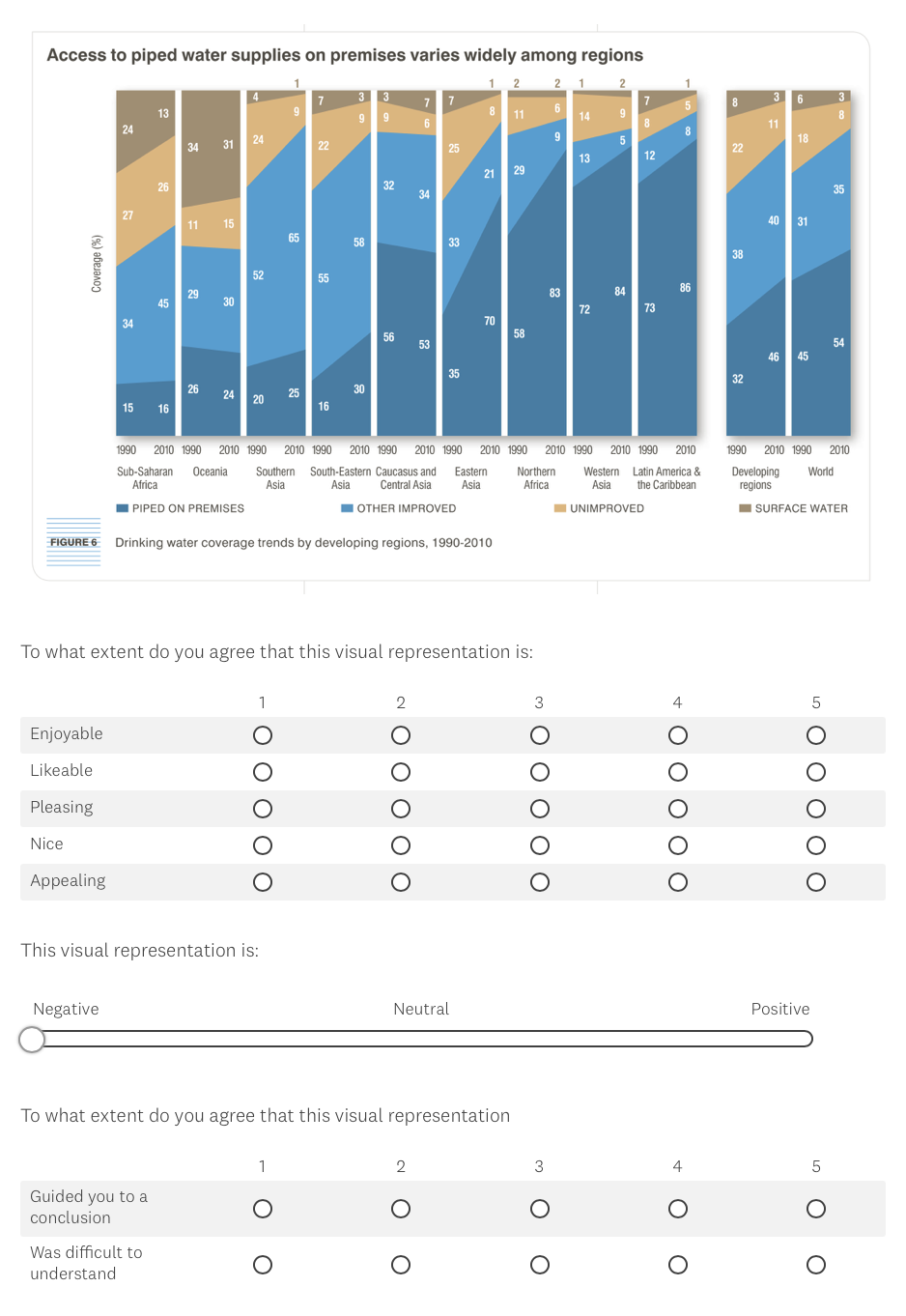}
  \label{fig:3b}}
  \caption{Description of tasks performed in Experiments 1 and 2.}
  \label{fig:3}
  \vspace{-1em}
\end{figure}

\textbf{Experiment 1:}\label{sec:expt1} As discussed in~\hyperref[sec:3]{Section 3}, we annotated 500 visualizations. We resized these while preserving aspect ratios to ensure their maximum dimension was 1000 pixels.  For each trial, participants were shown a visualization for 10 seconds. They were asked to perform a rating task where they positioned a slider on a bilinear scale ranging from -10 (Image) to +10 (Information), as shown in Fig.~\ref{fig:3a}. Each participant completed ten training trials, followed by 100 session trials. The 100 stimuli were randomly selected from the dataset of 500 visualizations, such that no chart appears twice during a session. An attention check appeared after the 25th, 50th, and 75th questions. Before running the main study, we conducted a pilot study with three participants to validate the design. We recruited 265 participants on Prolific\footnote{\url{https://www.prolific.co}} using the following user filters: (i) self-reported normal or corrected-to-normal vision, (ii) a first language of English, (iii) located in the U.S./U.K. with at least a high-school level of education, (iv) at least 100 previously completed Prolific surveys, and (v) an approval rate of 90\% or more.\footnote{\label{foot:demographics}See supplemental material for further demographic details.} Participants were paid $\$$3.00 for participation. Study duration averaged roughly 15:55 ($\pm$2:38) minutes, resulting in an average hourly pay of $\$$12.73 ($\pm$\$2.11). SurveyMonkey\footnote{\url{https://www.surveymonkey.com}} was used to display the study and store participant responses. We excluded 15 participants as their performance indicated they did not understand the task (i.e., more than 70\% of ratings had the values of 0, -10 or +10) during/after training, or they took more than 20 minutes to complete the study (>2$\sigma$). Each visualization was seen at least 50 times.

\textbf{Experiment 2:}\label{sec:expt2} We selected a subset of 100 stimuli from the dataset as exemplar visualizations. Of the exemplars, 11 are extreme examples of `image' ( $\overline{score}\in [-10,-6]$), 11 are extreme examples of `information' ( $\overline{score}\in [+6,+10]$), and the other 78 are in-between on the image-information spectrum, sampled equally across score intervals of $]-6,-3],]-3,3[,[3,6[$. Choosing them this way allows us to measure the effects of different visual elements while minimizing bias. The exemplar visualizations chosen exhibited the highest levels of inter-participant agreement within their respective score-interval, i.e., for each interval, charts with the lowest $\sigma$ for their average scores were picked (see Fig.~\ref{fig:5}). Thus, the exemplar population comprises the best-fitting visualizations for each interval and is representative of a real-world perceptual distribution. Participants perform two sets of tasks in succession:\footnote{\anj{All 100 stimuli are rated first; on completion, participants then perform the recall task on 50 randomly selected stimuli.}} (i) \textit{Rating}: This task set-up is identical to that used in \hyperref[sec:expt1]{Experiment 1}. (ii) \textit{Free Recall}: For each trial, participants viewed a stimulus for 30 seconds. Then, the screen was cleared, and participants were given 15 seconds to explicitly report aesthetic pleasure, perceived ease of use, and sentiment about the visualization. Finally, participants were asked to orally describe what they remember and feel about the visualization they viewed in the trial for 30 seconds. Each participant completed 50 trials, where each trial randomly selected one chart. Three training trials were performed before beginning the recall phase of the experiment. Similar to Experiment 1, a pilot study was first conducted to validate the design with three users. \anj{We recruited 101 participants from local communities belonging to two distinct groups. College-aged participants were recruited from a local university, and older participants were recruited from residential complexes in the locality via the leasing offices}.\footnote{Anonymized for the review stage} Our data was collected over a period of 4 months.\textsuperscript{\ref{foot:demographics}} We recruited 57 college-aged (27.2 $\pm$ 4.1 y/o) participants educated in STEM 
and 44 older participants  (57.5 $\pm$ 4.6 y/o) who are high-school (minimum) educated. All participants had normal color vision. Participants completed both phases of the experiment over a session lasting about 75 minutes. Participants volunteered their time in the study. The \textit{rating} and \textit{recall} phases were randomly ordered across participants in order to minimize effects of order bias. Each of the 100 visualizations was viewed at least once in the ratings task; 50 stimuli were seen for a second time in the recall task.

\subsection{Data Analysis}

\textbf{Ratings:}
We conducted a set of analyses for both experiments, comparing the rating scores to various visualization attributes (elicited based on prior research and the taxonomy (Table \ref{tab:1}) used to categorize the visualizations). The visualizations of the data analysis were constructed by summarizing across all visualizations. \anj{Our initial hypothesis was that different visualizations produce distinct internalizations. We further hypothesized that specific visual elements lead to diverging internalizations. However, as this work was exploratory, we did not have explicit predictions about which visual elements would produce various internalizations.} We performed regression analyses~\cite{hair2009multivariate,kefi2010using} and calculated Cohen's d to determine the effect strength of each factor (i.e., visual features)~\cite{cohen2013statistical} to preserve statistical power during analyses.\footnote{Detailed statistics are fully reported in the supplementary material \label{foot:results}} 

\anj{The results from the online study (Experiment 1) confirmed the existence of a distinct separation in the nature of visualization internalization and showed that there seemed to be some replicable variation in rating patterns across different age groups. However, since our online participants had high variation in (i) age, (ii) education, and (iii) each participant only rated a randomized subset of 100 stimuli from the full set of 500, we could not determine what proportion of noted patterns could be accounted for by individual differences due to small sample sizes. While formulating the online study, we noted that prior exploratory work that examined different aspects of the communicative power of visualizations had a high volume of results focused on surveying college-aged participants. One of the rare exceptions to this is Peck et al. \cite{10.1145/3290605.3300474}, who explicitly recruited a subset of participants that were older in age (they also considered these age brackets in their analyses). We follow a similar approach here, recruiting both college-aged students and also an older group of participants, to determine if age was a significant factor in rating patterns. 
}

\textbf{Recall:}
The explicit ratings for aesthetic pleasure, perceived ease of use, and sentiment were compared to the internalization scores using regression analyses. The free recall was analyzed using the engagement taxonomy proposed by Mahyar et al.~\cite{mahyar2015towards}. Higher levels of engagement include the formulation of a hypothesis based on the patterns and trends observed in the data. This hypothesis can be an initial assumption or idea about what the data may represent or indicate, which can then be further explored or tested. Subsequently, decision-making can be performed after validating the hypothesis. The formation of a hypothesis when viewing a visualization is often influenced by the viewer's prior knowledge and experience with the subject matter, as well as their cognitive biases and assumptions. Since participants viewed each stimulus for up to 30 seconds (recall) and also do not necessarily have domain expertise, we do not consider formulation and evaluation of hypotheses as contextual elements, as the quality of hypotheses has the potential to significantly vary between individuals.  

\begin{table}[H]
  \centering
\resizebox{0.7\columnwidth}{!}{%
\begin{tabular}{p{0.1\textwidth}p{0.35\textwidth}}
\toprule
\multicolumn{2}{l}{\textbf{Descriptive Elements}}                                                         \\ \midrule
\textbf{Text}       & Verbatim repetition of Title, Caption, Message                                      \\
\textbf{Annotation} & Arrows/Highlights/Text Labels                                                       \\
\textbf{Names}      & Using Text Labels/Legend to refer to specific Marks                                 \\
\textbf{Structure}  & Chart type, Skeuomorphism                                                            \\
\textbf{Data}       & Data type/source/domain                                                             \\
\textbf{Marks}      & Mentioning specific Data Points without reference to Name elements                  \\
\textbf{Channels}   & Features like Color, Shape, Size, etc.                                              \\
\textbf{Trends}     & Mentioning patterns observed in data without reference to Text or Name elements     \\
\textbf{Opinion}    & Likes/Dislikes of design choices (suitability of Structure, Mark, Channel elements) \\
\textbf{Sentiment}  & Emotional Response on viewing (e.g.: happy, sad, angry, positive, negative, etc.)  \\ \bottomrule
\end{tabular}%
}  
  \caption{Descriptive elements to parse in free recall verbiage.}
  \label{tab:3}
\end{table}

We therefore use the first three levels of Mahyar's taxonomy to determine the focus of free recall (see Table \ref{tab:3}): (i) Expose: the user can superficially understand what the visualization represents as a whole in terms of the source, domain, or topic of data. (ii) Involve (Interacting): the user can read and understand individual data points and visual encodings. (iii) Analyze (Finding Trends): the user can analyze the data to find trends, outliers, etc. We identified various descriptive elements to flag and parse during free recall, as shown in Table~\ref{tab:3}. \anj{The recall was manually transcribed during the study sessions; post study, 3 coders reviewed the transcripts and utterances were judged to identify which descriptive elements they could be mapped to. The presence or absence of these elements during recall, as well as the emphasis or order in which these elements are recalled, were noted.}

\section{Experimental Results and Discussion}

\subsection{Internalization Comparisons}

\begin{figure}
    \centering
    \includegraphics[width=0.4\textwidth]{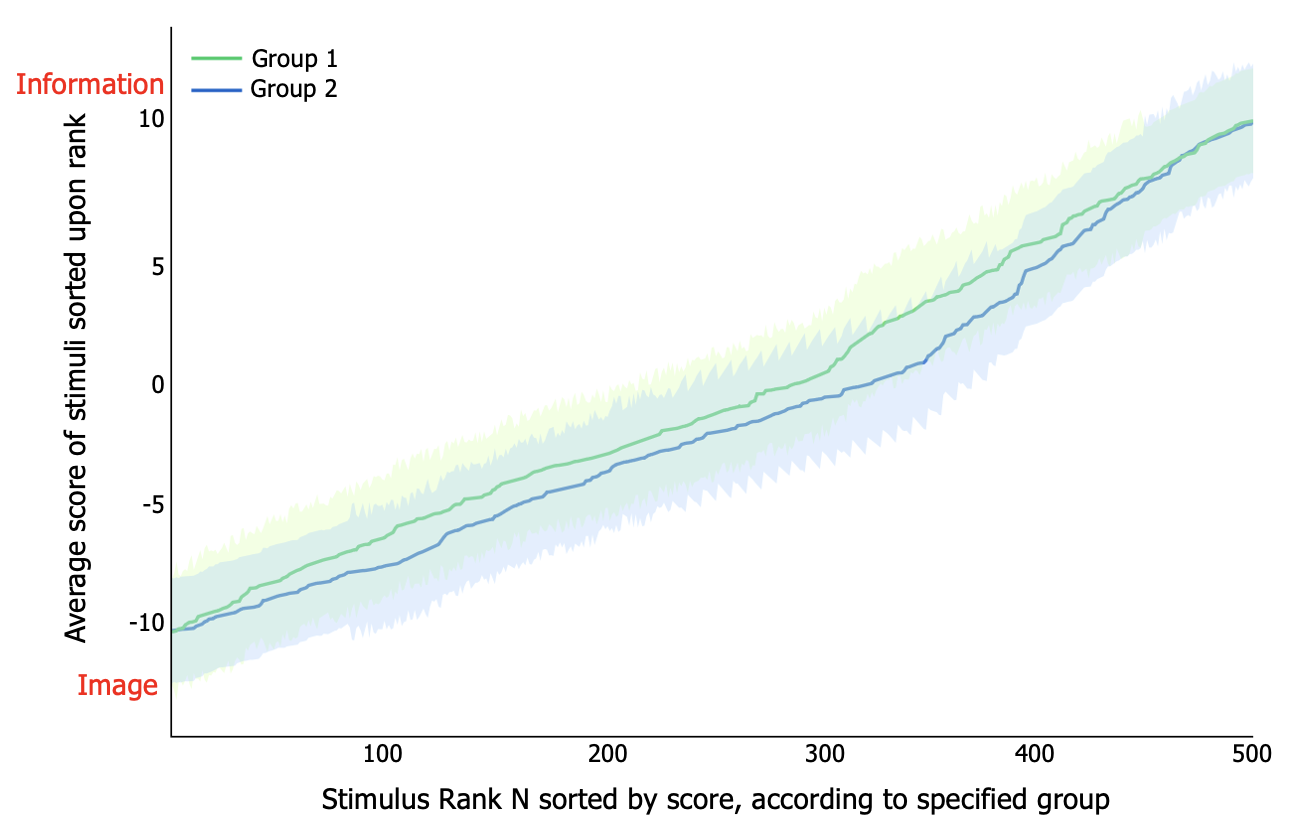}
    \caption{Participants were split into two independent sets, Group 1 and Group 2. Visualizations were ranked by internalization scores from participants in Group 1 (green line) or Group 2 (blue line) and plotted against the overall average internalization scores given by participants from both groups. Plots are averaged across 25 such random splits. Contours depict standard deviation averaged across ratings.}
    \label{fig:4}
    \vspace{-1em}
\end{figure}

Our experiments primarily present baseline results to identify the design elements that cause visualizations to be internalized in memory as `images' instead of `information'. We further examine how the nature of internalization  affects (i) the process of accessing and verbally recalling the visualization in the short term, (ii) aesthetic pleasure experienced on viewing the visualization, and (iii) the perceived ease of use for a visualization at-a-glance. We do not evaluate in the context of cognitive objectives that require longer viewing time, domain expertise, interactive charts, or long-term memorability. 
We measure the consistency of our internalization ratings by randomly splitting participants into two independent groups for comparison and averaging over 25 such half-splits to obtain a high Spearman’s correlation of 0.82 (see Fig.~\ref{fig:4}). 

\begin{figure}[H]
    \centering
    \includegraphics[width=0.35\textwidth]{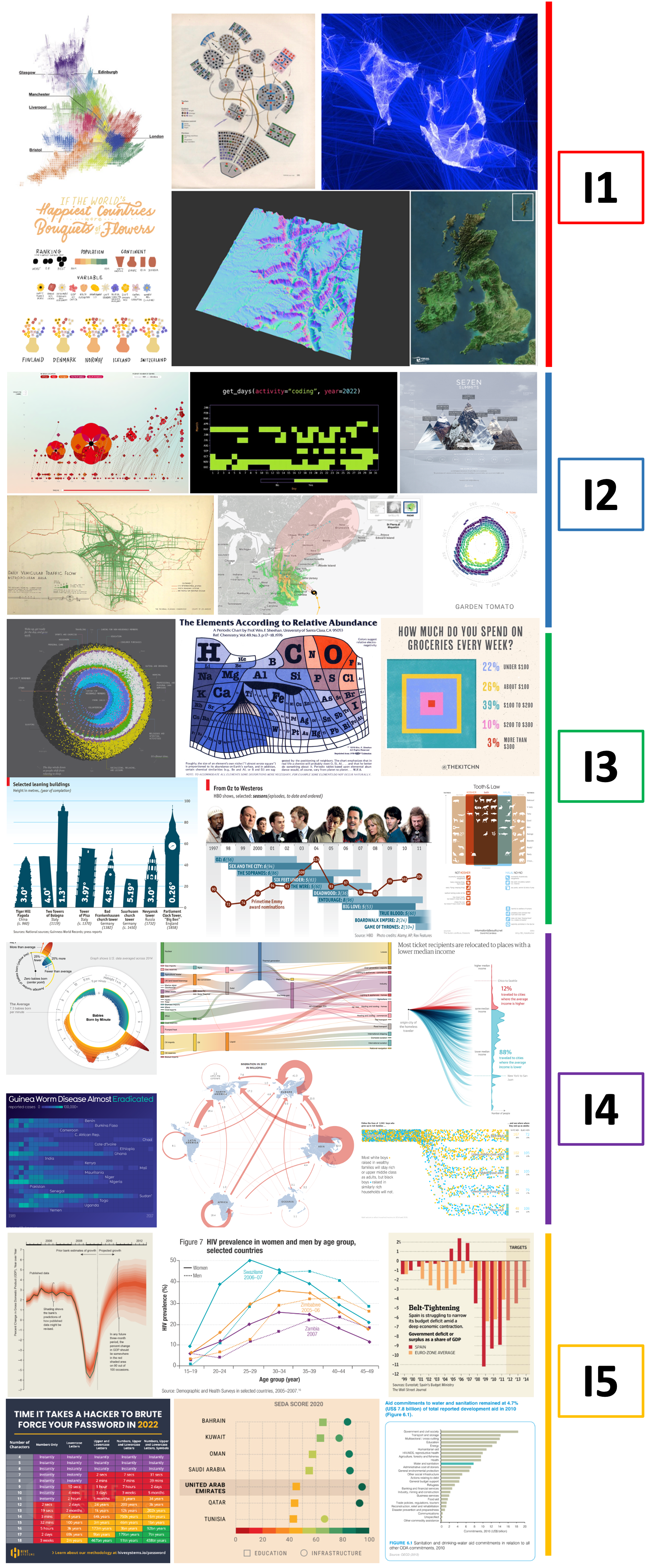}
    \caption{The top six visualizations with the highest inter-rater agreement for different score intervals. We organized the visualizations into groups (I1-I5) with I1 being those rated as the most image-like and I5 as the most information-like.}
    \label{fig:5}
    \vspace{-1em}
\end{figure}

We represent 30 visualizations with the lowest $\sigma$ (i.e., highest rating agreement) across various score-intervals (\textbf{I1}--\textbf{I5}) in Fig.~\ref{fig:5}. Overall, there is a strong agreement in chart ratings: $\sigma \in [2--2.73]$. Additionally, in Experiment 2,  we find that older participants  exhibit higher inter-rater agreement (mean $\sigma=2.25$) across all score intervals for our subset of 100 visualizations, in comparison to younger participants (mean $\sigma=2.68$). \anj{We also compare the ratings between Experiments 1 and 2 for the selected stimuli subset and find that in Experiment 2, the stimuli fall in the same scoring intervals and have ratings within 0--2 points of that in Experiment 1.}

\subsection{The Effects of Visualization Elements on the Nature of Internalization}

\begin{figure*}[!ht]
    \centering
\begin{subfigure}{.4\textwidth}
  \centering
  \includegraphics[width=\linewidth]{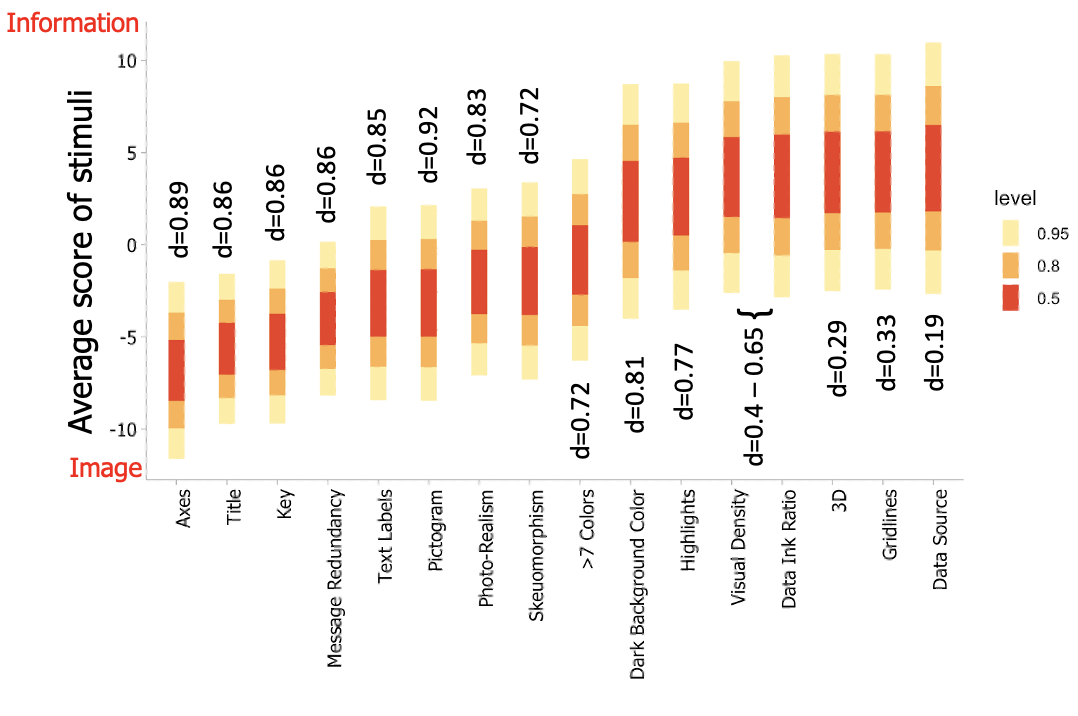}
  \caption{Absence/Low-Level}
  \label{fig:sub1}
\end{subfigure}%
\begin{subfigure}{.4\textwidth}
  \centering
  \includegraphics[width=\linewidth]{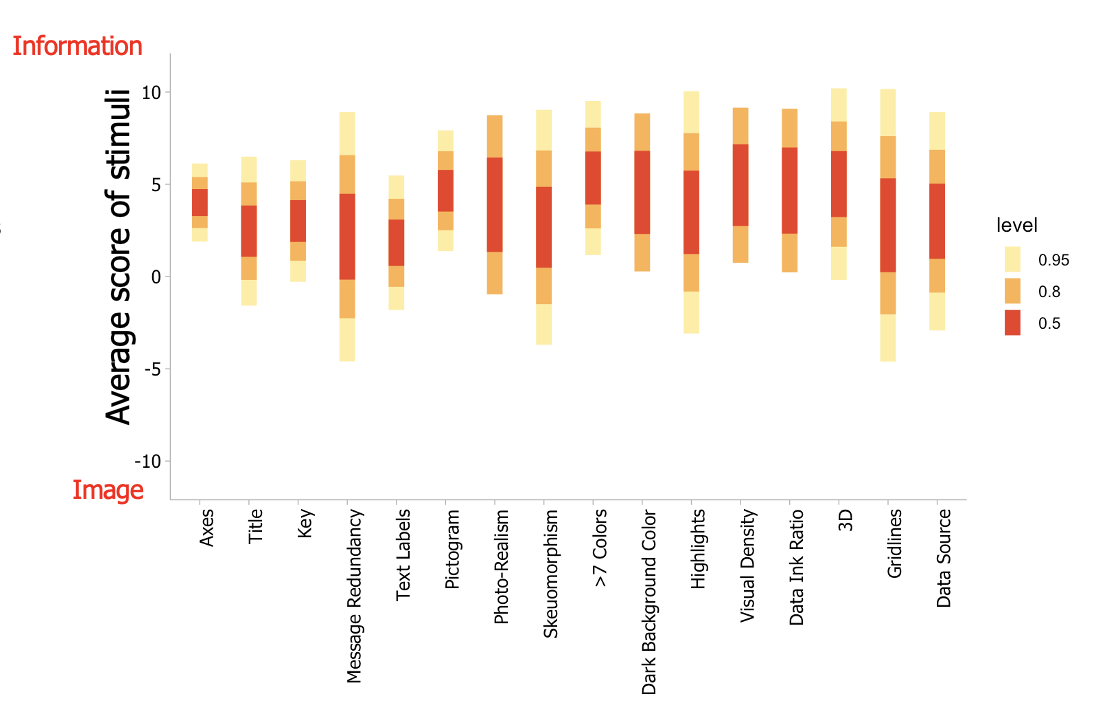}
  \caption{Presence/High-Level}
  \label{fig:sub2}
\end{subfigure}
\caption{Internalization scores averaged for visualizations with different visual elements. Each element is considered in isolation for its reported score, with intervals depicting the score distribution \anj{and effect size denoted using cohen's d values.}}
    \label{fig:7}
    \vspace{-1em}
\end{figure*}

\begin{figure}
    \centering
    \includegraphics[width=0.9\columnwidth]{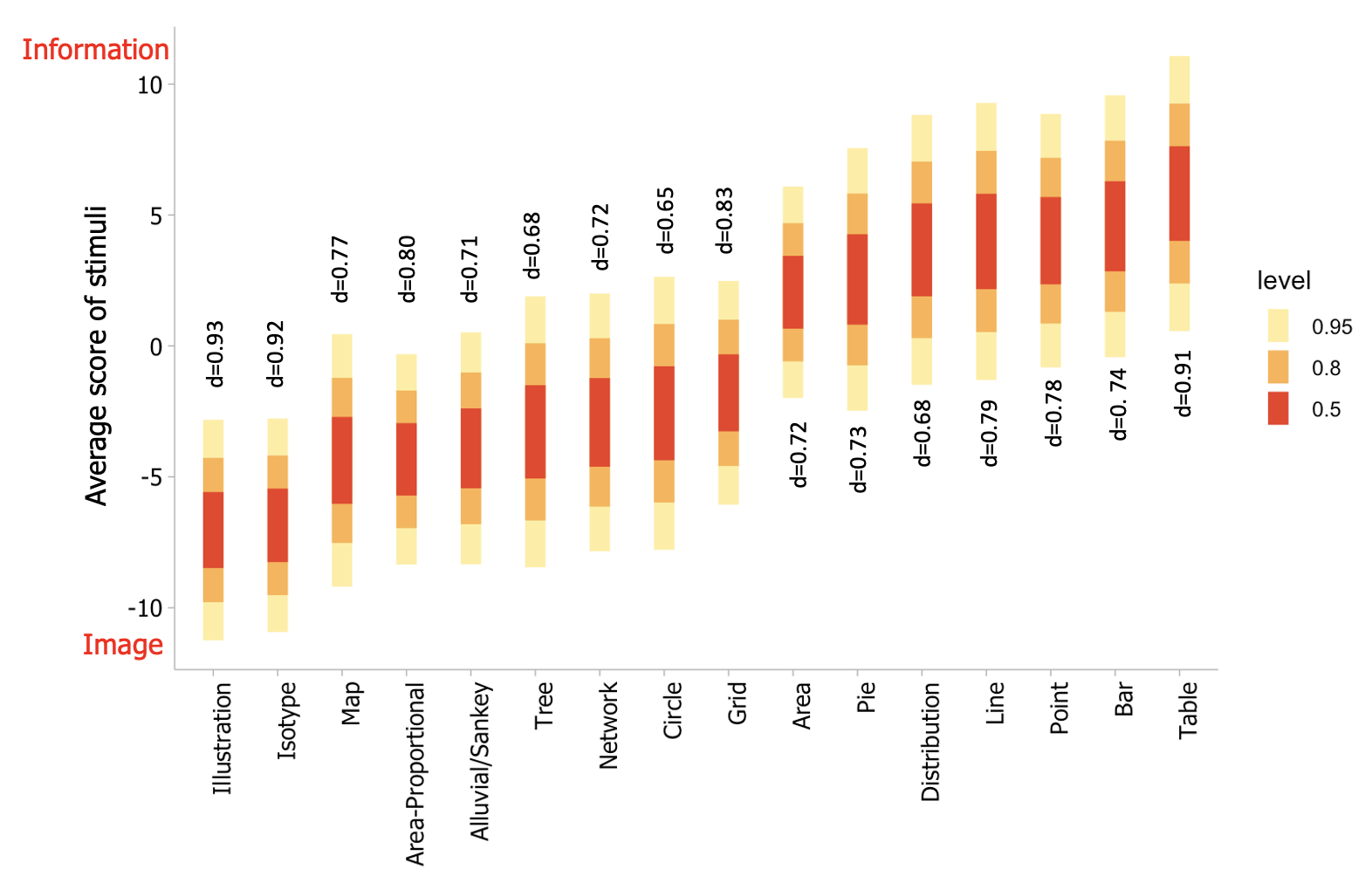}
    \caption{Internalization scores averaged for visualizations based on visualization type. Note that a category is taken to comprise of all the chart sub-types other than the ones that have been explicitly separated with intervals depicting the score distribution \anj{and effect size denoted using cohen's d values.}}
    \label{fig:6}
    \vspace{-2em}
\end{figure}

We consider the individual impact of each design element\footnote{\anj{The symbols} $\textcolor{green}{\blacktriangle}$/$\textcolor{red}{\blacktriangledown}$ \anj{denote increase/decrease in informational value of stimuli based on element presence.}} in our discussion below and represent average internalization scores in Figs.~\ref{fig:7} and~\ref{fig:6}. A common trend observed in our analysis is that older participants display larger effect sizes ($d$ values are $0.09\pm0.03$ greater on average) for individual factors, that lead them to rate visualizations in a more consistent manner when compared to younger participants.

\textbf{Chart-type, Skeuomorphism, Human Recognizable Objects, Human Depiction ($\textcolor{red}{\blacktriangledown}$) :} The internalization scores for different chart types are summarized in Fig.~\ref{fig:6}. Of our 500 target visualizations, we observe that 283 contained either skeuomorphism, photographs, or pictograms of human recognizable objects (including human depiction). Visualizations containing pictograms are on average considered more \textit{image-like} (internalization score -6.85$\pm$2.08\footnote{\label{sig}All results discussed are statistically significant, with $p<0.001$}) than those without pictograms (internalization score 3.22$\pm$3.01). This supports the observation that isotype charts and chernoff diagrams are in the more image-like categories (\textbf{I1} and \textbf{I2} in Fig.~\ref{fig:6}) (R$^2$=0.84, d=0.92). Charts with photo-realistic glyphs (such as the usage of real-world objects or human faces) also follow this trend (R$^2$=0.82, d=0.83). Area charts that rely on proportional comparison without axes also follow this trend and are found in \textbf{I1} and \textbf{I2} (R$^2$=0.79, d=0.80); the individual data elements can be regarded as variably sized pictograms in these visualizations.

Surprisingly, we observe that cartographic representations also feature heavily in \textbf{I1} and \textbf{I2} (R$^2$=0.76, d=0.77), particularly for maps that do not have overlaid text annotation (R$^2$=0.80, d=0.81). A potential explanation for this might be that geographical landmasses are skeuomorphic in nature, and can be treated akin to pictograms. Indeed, on further analysis, we note that skeuomorphic charts are perceived as more image-like (R$^2$=0.71, d=0.72). Radial warping (such as in chord diagrams, donut charts, timelines, radial bar charts, etc.) also leads to ratings between \textbf{I1} and \textbf{I2}, though with lower effect size (R$^2$=0.72, d=0.65). Pie charts, though radial in appearance, do not exhibit such an effect. We posit that the presence of curvature may influence this internalization of relatively unfamiliar, visually complex representations, though further investigation is needed to determine the exact cause. 

In terms of \textit{information} ratings, on average we find that bar$>$point$>$line$>$pie$>$area charts; these visualizations are also perceived as more informative than all other visualization types in the study. However, scores exhibit wide, seemingly random variation between \textbf{I3}--\textbf{I5} when considering chart-type as the only determining factor. We attribute increased \textit{information} ratings to participant familiarity with such charts, with other design elements' presence/absence dictating the sub-categorization of these charts amongst score intervals.

\textbf{Presence of Axes ($\textcolor{green}{\blacktriangle}$) :} There is an observable trend of visualizations with conventional axes representation (i.e., the presence of linear x and y-axes) being rated in \textbf{I3}--\textbf{I5} (R$^2$=0.87, d=0.89). The presence of tick marks and axes labels increases the scores to \textbf{I4}--\textbf{I5} (R$^2$=0.86, d=0.91) across different chart types, data domains, data scales, visual density, and data-ink ratio. Given that participants have a high familiarity with and exposure to visualizations like point (scatter/bubble-plots), line, bar, and area charts in the real world, we believe that the presence of axes intrinsically boosts the perception of stimuli as more `informative.'

\textbf{Titles and Keys ($\textcolor{green}{\blacktriangle}$) :} When participants were shown stimuli with titles during the rating task, they found those visualizations to be relatively more informative (both within and across all five score intervals), than those lacking these titles (R$^2$=0.83, d=0.86). However, for charts with glyphs representing more than one dimension, the combined presence of a title and a key exhibit higher \textit{information} effect size (R$^2$=0.78, d=0.73). In the case of charts with only a key present, the \textit{information} rating is boosted, but shows a smaller effect size than titled charts (R$^2$=0.77, d=0.69). These observations align with prior literature that highlights `titles' as high-priority design elements that draw visual attention and increase the perceived quality of visualizations~\cite{borkin2013makes,wanzer2021role}.

\textbf{Annotation ($\textcolor{green}{\blacktriangle}$) :} We observe that visualizations with high levels of text annotation are scored as highly \textit{informative} $\in$ \textbf{I5} (R$^2$=0.92, d=0.85). The presence of message redundancy (R$^2$=0.79, d=0.86) and mark labels (R$^2$=0.74, d=0.83) also significantly bias \textit{information} internalization. On the other hand, the use of highlighting and overlay of glyphs (like arrows) for annotation show relatively smaller effect sizes (R$^2$=0.77, d=0.68 and R$^2$=0.72, d=0.65, respectively). We attribute this to an increased preference for charts annotated with more text than sparser counterparts, in line with previous work~\cite{9904452}. Additionally, the text explicitly describes statistical or relational components of a chart (or even summarizes chart takeaways through message redundancy) in comparison to highlights or arrows, which tend to focus on elemental or encoded components. We therefore speculate that participants perceive charts that explicitly state the output of higher-level cognitive tasks such as summarization or decision making as `easy to use' and therefore internalize them as \textit{information}.

\textbf{Usage of Color and 3D ($\textcolor{red}{\blacktriangledown}$) :} More colorful visualizations exhibit \textit{image} internalization: visualizations with seven or more colors display the largest effect size (R$^2$=0.73, d=0.72), followed by those with 2--6 colors (R$^2$=0.71, d=0.70), and finally visualizations with one color or black-and-white gradient (R$^2$=0.65, d=0.68). We also analyze visualization scores based on a background color; stimuli with white backgrounds are more likely to be rated as \textit{information} (R$^2$=0.83, d=0.79), while non-white backgrounds promote \textit{image} ratings. The effect of the background color is more pronounced for visualizations with darker hues (R$^2$=0.86, d=0.81). Additionally, the presence of gridlines and usage of 3D glyphs show a relatively low impact on internalization ratings (with d=0.33 and d=0.29, respectively). Hence, we do not further analyze internalization scores on these bases. 

\textbf{Data Ink Ratio, Visual Density ($\textcolor{green}{\blacktriangle}$) :} Considering all the visualizations together, the respective influences of data-ink ratio and visual density on internalization have moderate effect sizes for the above results (d $\in$ 0.4--0.65). Low data-ink ratio and high visual density bias ratings towards \textbf{I1}--\textbf{I3}. Additionally, moderate--high data-ink ratio and low--moderate visual density bias ratings towards \textbf{I3}--\textbf{I5}. The broader range of variation in rating scores indicates that these factors, while significant, are not of high priority during internalization.

\textbf{Data Source and Domain ($\textcolor{green}{\blacktriangle}$) :} Some visualizations display explicit information about the `source' and `domain' of data via text annotation, typically located bottom-left/right of the stimuli. These elements display negligible effect sizes (d$\leq$0.19) and therefore do not significantly influence chart internalization.

\textbf{Data Scale and Dimensionality ($\textcolor{red}{\blacktriangledown}$) :} We observe that the scale and dimensionality of data serve as moderators for all the attributes discussed above while displaying negligible effect sizes (d$\leq$0.15) when considered in isolation. Moderate--high dimensionality and moderate--high data scale lead to the relatively increased internalization of visualizations as \textit{images} overall attributes; lower values of the two do not significantly impact ratings. We can attribute this phenomenon to an increase in the perceived complexity of charts with larger data volumes, which can overwhelm viewers, decreasing their efficiency in distilling relevant information from the chart. 

\textbf{Synthesizing these Results: } \anj{We note that the presence of elements does not show a clear opposite trend to absence of elements in Fig. \ref{fig:7}; we attribute this to these elements being considered in isolation in our analysis. For instance, a chart in I5 might comprise axes, title, key,  and message, and these factors might in combination produce an average rating>5. We accordingly synthesize} a rudimentary ranking of the influence various design elements have on the internalization of a visualization: \textcolor{red}{axes}$>$\textcolor{red}{titles/keys}$>$\textcolor{red}{message redundancy annotation}$>$\textcolor{red}{text label annotation}$>$\textcolor{teal}{pictogram/photo-realistic glyphs}$>$\textcolor{teal}{more than 7 colors/dark background-color}$>$\textcolor{red}{highlight or glyph-based annotation}$>$\textcolor{teal}{visual density}$>$\textcolor{teal}{data-ink ratio}$>$\textcolor{teal}{3D/gridlines}$>$\textcolor{teal}{data source/domain}. Here, attributes are ordered such that their \textcolor{teal}{presence/high-level} or \textcolor{red}{absence/low-level} increases the likelihood that the visualization is internalized as an \textit{image}. Data scale and dimensionality levels increase the effect sizes for this ranking. By reversing the \textcolor{teal}{presence}/\textcolor{red}{absence} of these design elements in the ranking, we can expect an increased likelihood that the visualization considered will be internalized as \textit{information}. 

\subsection{``At-a-glance'' vs. ``Prolonged Exposure'' Internalization}

\begin{figure}[H]
    \centering
\includegraphics[width=0.4\textwidth]{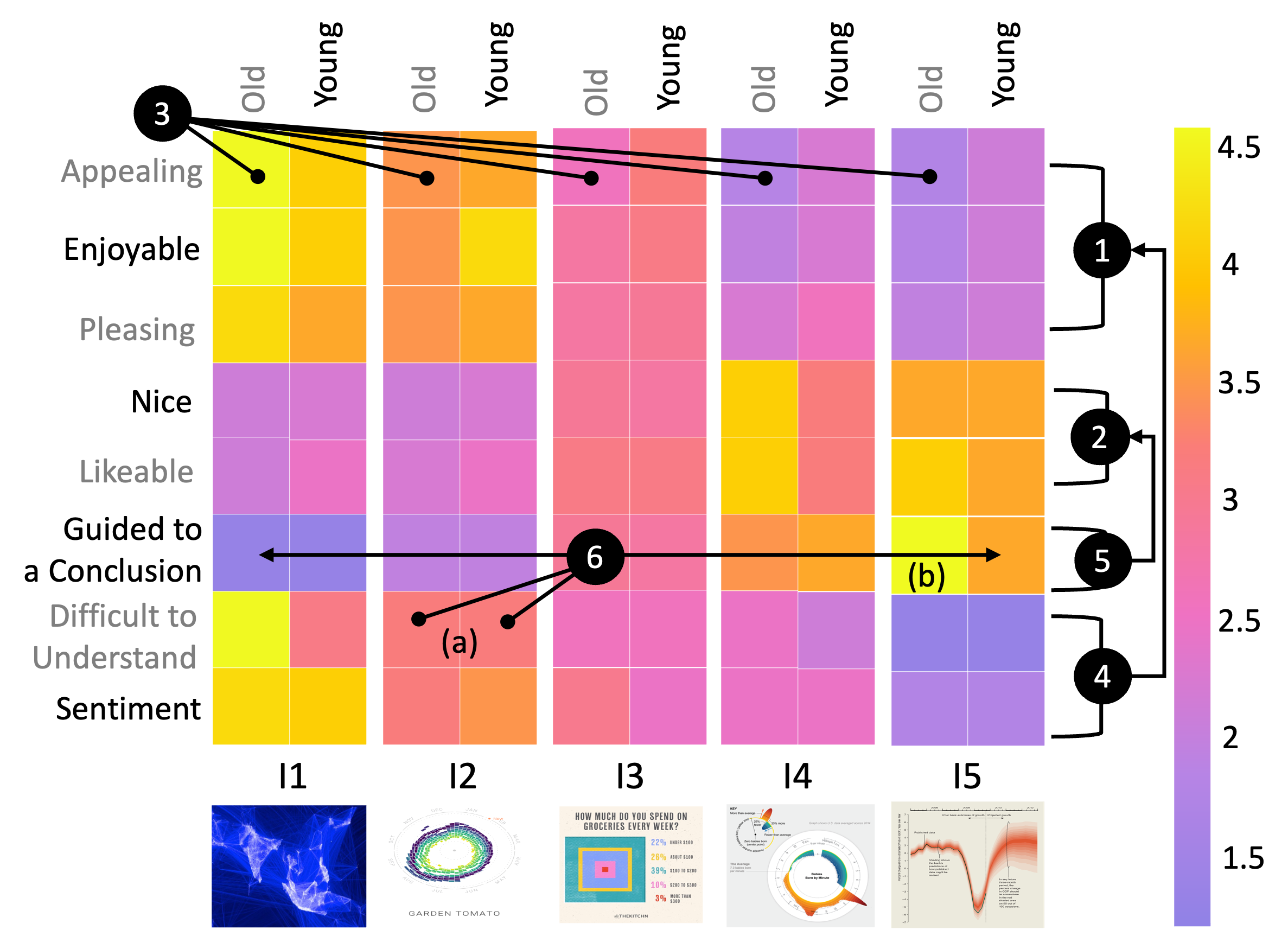}
    \caption{Internalization score intervals for visualizations juxtaposed against aesthetic pleasure, perceived ease of use, and viewer sentiment across different user groups. We consider sentiment ratings of positive/neutral/negative to correspond to scores of 5/3/1. \anj{(1) \textit{Emotion-Aesthetics} scores are higher when stimuli are perceived as more image-like. (2) \textit{Data-Aesthetics} scores are higher when stimuli are perceived as more informational. (3) Older participants show greater variation in Emotion-Aesthetics ratings between I1--I5. (4) \textit{Sentiment} and \textit{Perceived Ease of Use} are proportional to Emotion-Aesthetics scores. (5) \textit{Feeling Guided} is proportional to Data-Aesthetics scores. \textit{Feeling Guided} ratings: (6a) display the most equivalence between the two participant groups. (6b) show the highest range of variation between I1--I5.}}
    \label{fig:8}
\end{figure}

We restrict our analysis of data collected during free recall by juxtaposing only against \textbf{I1}--\textbf{I5} for rated stimuli (see Fig.~\ref{fig:8} for rating patterns and high-level insights). This constraint follows from the non-standardization of dimensionality and data volume across visualizations, which significantly impact the nature of element-wise design choices made during visualization creation.

\textbf{Aesthetic Pleasure:} 
\anj{The Beauvis scale~\cite{he2022beauvis} assesses a visualization's aesthetic pleasure (or beauty). The authors use the terms `appealing,' `enjoyable,' `likable,' `nice,' and `pleasing' to target \textit{emotion-oriented} or affective judgment of a visualization; this scale does not target \textit{data-aesthetic} judgment, which hinges on the combination of data and design.} We find that \textit{image} visualizations $\in$ \textbf{I1}--\textbf{I2} elicit higher ratings for being `appealing' (R$^2$=0.82, d=0.73), `enjoyable' (R$^2$=0.85, d=0.74), and `pleasing' (R$^2$=0.77, d=0.73). However, such visualizations receive lower scores for `likable' (R$^2$=0.74, d=0.71) and `nice' (R$^2$=0.68, d=0.70). \anj{We conjecture that `nice' and `likable' are partially conflated as a data-aesthetic judgment and are not grouped with the other emotion-oriented terms, as observed in Figure~\ref{fig:8}(2).} On the other hand, visualizations rated as \textit{information} $\in$ \textbf{I3}--\textbf{I5} elicit higher ratings for being more aesthetically `likable' (R$^2$=0.73, d=0.65) and `nice' (R$^2$=0.69, d=0.67), and low--moderate ratings for being `appealing' (R$^2$=0.76, d=0.71), `enjoyable' (R$^2$=0.76, d=0.68) and `pleasing' (R$^2$=0.74, d=0.69). \anj{This could be attributed to the manner of defining ‘images’ within the scope of our study: we associated images with invoking a sensory reaction which did not necessarily evoke analytical thought. Based on this definition, participants seem to have associated ‘appealing’, ‘enjoyable’, and ‘pleasing’ with such a sensory experience.} Older participants show  clearer patterns of high emotion-oriented aesthetic enjoyment of \textit{image}-rated visualizations ($d$ values are $0.07\pm0.04$ greater on average) compared to younger subjects; in the case of data-aesthetic judgments, the effect size is smaller across both groups of subjects, though it remains significant, as shown in Figure~\ref{fig:8}(3).

\textbf{Perceived ease of use and sentiment:} Participants feel less `guided to a conclusion' by \textit{image} visualizations $\in$ \textbf{I1}--\textbf{I2} (R$^2$=0.78, d=0.68) and express higher \textit{positive} sentiment (R$^2$=0.69, d=0.70) when viewing these stimuli. However, they simultaneously rate such visualizations as comparatively `difficult to understand' (R$^2$=0.73, d=0.71). \anj{Feeling `guided' is strongly associated with data-aesthetics measures (Figure~\ref{fig:8}(5)); it is also the measure for which both groups of participants have the most equivalent rating (Figure~\ref{fig:8}(6a)) across I1-I5, which can be potentially attributed to the lower volumes of text present in image-like stimuli. The absence of text places a higher internal cognitive demand on a viewer to process the information shown and form a useful takeaway. The presence of elements like title/annotation/message/etc. in the visualizations causes the cognitive demand on participants to decrease rapidly, in turn reflected in the ratings for `feeling guided' to fluctuate to a greater extent (Figure~\ref{fig:8}(6b)) compared to other measures. } \textit{Information} visualizations $\in$ \textbf{I4}--\textbf{I5} are considered less complex and provide explicit guidance but are associated with more neutral/negative sentiments. \anj{Sentiment hence is associated by participants with emotion-oriented aesthetic judgements (Figure~\ref{fig:8}(4)), and ease of use follows a similar trend.} This confound indicates a dissociation between the \textit{pleasure derived} when viewing a visualization versus the \textit{perceived communicative utility} of the visualization. 

\begin{figure}[!ht]
    \centering
  \includegraphics[width=0.3\textwidth]{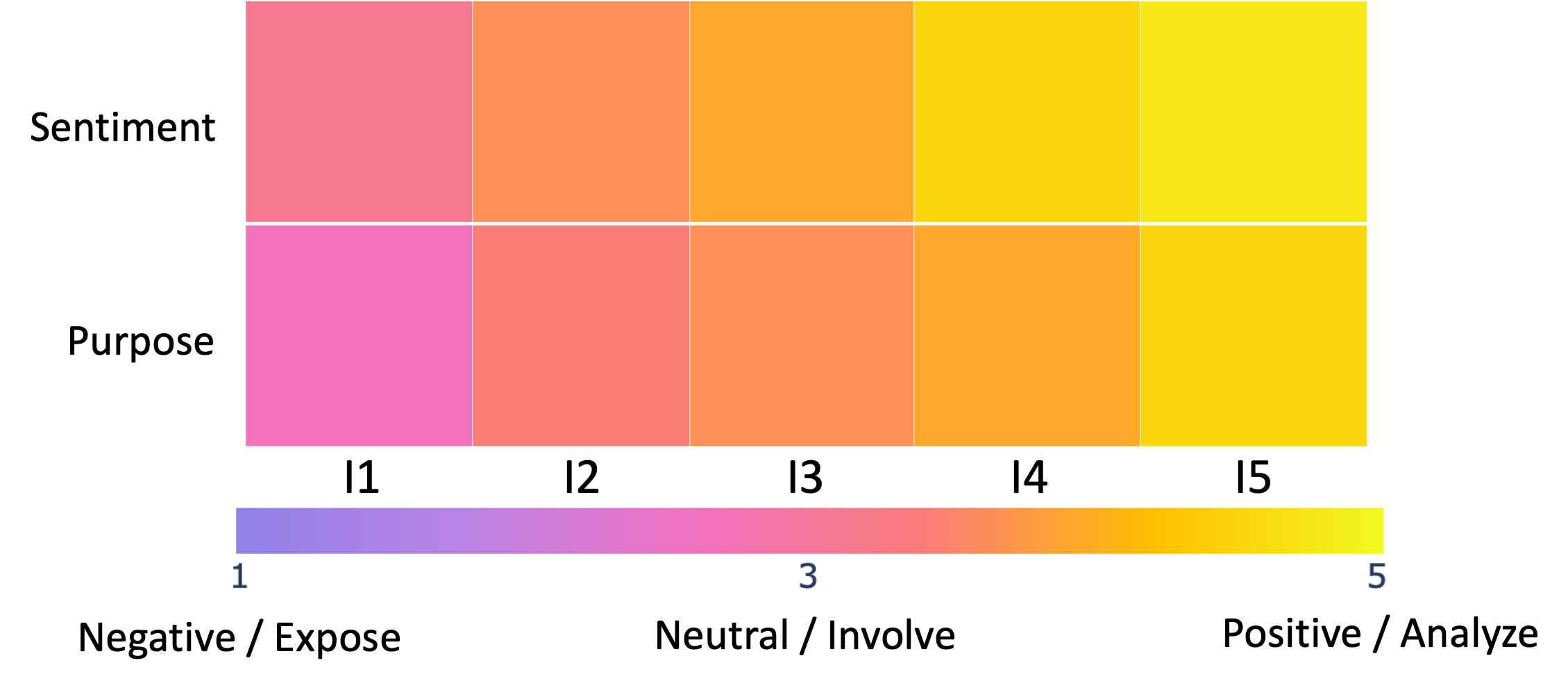}
    \caption{Internalization score intervals compared to expert annotations for sentiment and purpose, rated as 5/3/1, with 5 corresponding to more positive and higher engagement levels, respectively. }
    \label{fig:9}
\end{figure}

\textbf{Designer Intent:} As mentioned in Table~\ref{tab:1}, visualizations were annotated by experts based on their \textit{sentiment} (positive/negative/neutral emotional response) and \textit{purpose} (expose/involve/analyze). The experts annotated the visualizations by considering them from a designer's perspective; they estimated how a viewer would respond emotionally to these visuals and how engaging they would find them. We found that experts coded \textbf{I4}--\textbf{I5} visualizations as \textit{positive}, compared to other intervals (R$^2$=0.82, d=0.78), as shown in Fig.~\ref{fig:9}. Additionally, these intervals were associated with viewer intent to \textit{analyze} (R$^2$=0.85, d=0.80). \textbf{I3} visualizations are tagged as \textit{neutral}, and \textbf{I1}--\textbf{I2} as \textit{neutral/negative}. Experts, however, annotated very few visualizations from \textbf{I1}--\textbf{I3} (<5\%) with the \textit{expose} level of engagement; it is only used when there is no text present. Instead, \textit{involve} is prevalent for \textit{low} text volume. Experts, therefore, seem to associate more \textit{positive} sentiment with visualizations seen as more \textit{informative}. \anj{This however, contradicts participant sentiment ratings as shown in Fig.~\ref{fig:8}, wherein more informative stimuli were rated as having neutral/negative sentiment.}

\textbf{Verbal Recall Patterns:}
Our analysis of verbal recall patterns can be summarized as follows: \textit{Image}-rated visualizations focus (in priority order) on the recall of (i) individual data points (often those that are either annotated with a label/highlight, or have max-values), (ii) description of visual encoding appearance (typically in order of type of axes, human-recognizable glyphs, color, shape), (iii) positive or neutral opinion on the aesthetic appeal of encoding choices, and (iv)~stating the data domain and reporting moderate--high difficulty of understanding data nuances. Such charts, therefore, seek to \textit{expose} and, in some cases, \textit{involve} the user in performing low-level tasks related to data interpretation. \textit{Information}-rated visualizations focus (in priority order) on the recall of (i) the title/message of the visualization, (ii) text-label annotations on marks and axes labels, (iii) reporting high ease of use, observed trend, and re-stating the title/message \anj{(paraphrased and in context of an observed trend or annotation)} of the visualization, (iv) stating the presence of highlight or overlay annotations like arrows and chart-type (if known). Such charts, therefore, seek to \textit{involve} the user in performing tasks to \textit{analyze} data trends and generate chart takeaways.

\anj{Notably, recall patterns align with the expert view of a visualization's purpose (as discussed above); experts rank more informative stimuli as serving analytical purposes (`analyze' level), while image-like stimuli are attributed to being designed for creating awareness (`expose' level) on the topic of the visualization to viewers. Accordingly, recall patterns for more informative stimuli (I4--I5) focus on stating a trend or takeaway.  Stimuli in I2--I3 tended to have a sufficient amount of annotation for a viewer to be involved in recalling particular data points of interest, in accordance with the intermediate `involve' level of Mahyar's taxonomy. Image-like stimuli (I1--I2) on the other hand, are primarily described in terms of their appearance, with statements about the domain (or topic) of the visualization taking low priority primarily via referencing.} 

To summarize, visualizations perceived as `images' may be less efficacious for communicating trends and messages, although they tend to evoke a more positive emotional response. On the other hand, 'informative' visualizations focus on annotation-based recall and receive more positive design evaluations.

\section{Conclusions and Future Work}

\anj{
Our research has shed light on how people internalize and recall visualizations, and  provides strong quantitative evidence supporting traditional guidelines for designing and presenting visualizations. We have found that certain design elements can affect how viewers pay attention to and remember visualizations. These elements can serve as effective cues for recalling and guiding viewers through the visualization. Particularly, we find that elements such as axes, title, key, high text volume, etc., minimalist representations, and more familiar chart types (such as bar, pie, line, etc.) are viewed as contributing to stimuli having high perceived informational value. We speculate this result arises from the overall appearance of such stimuli being closer to the abstract canonical impression of particular visualization types. Furthermore, we observe a distinct trend of older participants exhibiting more uniform patterns of rating and recall, with larger effect sizes and lower standard deviation. A potentially stronger dependency (arising from accumulated knowledge and exposure) on `canonical' representations, combined with potential declines in memory due to aging~\cite{tas2020age,leonards2002role} might partially account for this phenomenon, though it merits deeper inquiry.

We note that we frame our constructed definitions of ‘image’ vs. ‘information’ such that study participants consider an image to evoke a primarily superficial sensory reaction that afforded the development of little to no analytical insights, although a viewer might form strong aesthetic-based judgements. We chose a bilinear scale in accordance with this view, as a visualization can have high aesthetic impact, while leading to varying levels and quality of analytical insight~\cite{borkin2015beyond}. This primarily arose from a lack of vocabulary to define something with purely sensory impact and no informational value in terms of visuals. Additionally, our definitions consider design, purpose, and response, though purpose and design are often studied as a cohesive unit~\cite{10.2312:evs.20211061}. However, while a visualization's design might be suitable for a viewer to elicit an accurate takeaway, the takeaway generated might be of low value; for instance slow analytics~\cite{bradley2021approaching} prioritizes retention and understanding as the higher purpose of a visualization, while 'precision' based takeaways are of lower value.  As this work is exploratory, future studies can better study the competing and nuanced considerations of image and information. 
}

The following conclusions summarize our high-level insights:

\textbf{\textit{Information} elicits positive design judgments and negative emotional judgments.}
Visualizations with high communicative utility likely require less cognitive load in discerning visualization takeaways. Importantly, when these visualizations are retrieved from memory, many annotation-related details of the visualization are also retrieved. Thus, participant-generated descriptions tend to be of higher quality for these visualizations. However, such visualizations simultaneously evoke a low-level, negative-leaning emotional response, indicating a confound between designing for affective vs. cognitive objectives.

\textbf{\textit{Images} elicit positive emotional judgment and are regarded as more visually complex.}
Visualizations regarded as aesthetically pleasing are perceived to have low analytical utility. This is observed even in the case of pictograms, which have been previously shown to serve as effective visual hooks into memory~\cite{borkin2015beyond}, allowing visualizations to be better recognized and described. The nature of description, however, supports this view of increased `complexity,' as descriptions tend to center around the appearance of visual encodings or individual data points. Users simultaneously exhibit low analytical engagement and high emotional engagement with a given visualization.

\textbf{Redundancy improves quality of both design and emotional response.}
When a visualization contains redundant information, such as repeating quantitative values (data redundancy), the audience tends to make more emotional/affective judgments. On the other hand, if the redundancy covers the main trends or concepts of the visualization (message redundancy), the audience tends to make more design-related judgments. It is essential to strike a balance between the levels of data and message redundancy to provide clarity and organization to complex information, for instance, by manipulating the relative positions of different text annotations with respect to visual encodings used. This balance can also facilitate comparisons across different parts of the data, resulting in a visually appealing and balanced design.

\textbf{Embellishing familiar charts can preserve the perceived communicative utility and increase emotional response.}
A stylistic embellishment of common visualization types, such as bar charts or bubble charts, can partially bridge the gap between emotion and design-based judgments. Participant familiarity with such visualization types may be able to offset the detrimental impact of increased visual density from embellishments on perceived design quality; \anj{further examination of the relationship between perceived vs. encoded information is necessary to develop more concrete guidelines on what is ``useful" communicative aesthetic.} A similar observation holds for cartographic visualizations minimally overlaid with text annotations. \anj{Additionally, the dichotomy of our rating task between image and information necessitates that users associate 'images' with evoking a primarily superficial sensory reaction as opposed to cognitive processing. Detailed examination of stimuli in \textbf{I3} might provide further insights on balancing functional (\textit{information}) and aesthetic design (\textit{image)} considerations.}

Simply understanding how the brain processes visualizations is the initial stage towards comprehending how to effectively communicate the crucial and pertinent aspects of the data or trend that the designer intends to convey in a way that ensures the viewer retains them. To gain further insights into visualization internalization, we plan to expand our current database, particularly considering that  only a best-fit subset of our current dataset is studied in the context of recall. 

\anj{Additionally, while participants rate the aesthetics, sentiment, and ease of use (Fig.~\ref{fig:9}) prior to verbal recall, we did not analyze the recall against each of these individual ratings. We do observe broad general trends between recall and ratings; for instance, for a stimuli in I1, the ratings are high for `aesthetics' and `sentiment', low for `data-aesthetics' and `ease of use', and verbal recall primarily mentions individual annotated data points or describes visual encoding appearance. It's possible that rating prior to recall helps participants organize their thoughts for recall, enabling uniform qualitative analysis by coding for particular aspects of recall. In the future, we intend to code and analyze recall data as well as} use more fine-grained definitions and measures of abstract concepts like visual density and data scale to better understand specific subtleties in various sub-types of visualizations. We also plan to investigate the impact of time on the quality of visualization recall by conducting experiments based on eye-tracking, change-detection, short-term and long-term memorability, and determining which visual elements stick in viewers' minds longer. Furthermore, we aim to directly demonstrate in future work how the nature of internalization significantly affects visualization comprehension and how this may lead to mind-wandering during visualization viewing and recollection. Lastly, we plan to design and evaluate interaction techniques that can help intrinsically resolve user uncertainty when exploring image-like visualizations to boost analytic engagement while maintaining emotional engagement.

By gaining a better understanding of how visualizations are internalized in memory, future studies can ask more high-stakes questions about what makes a visualization effective or engaging. Understanding the impact of low-level visual elements on internalization can help control for them in future experiments, ultimately leading to guidelines to creatively engineer high-impact holistic visualizations.

\section*{Supplemental Materials}
\label{sec:supplemental_materials}

Refer to the following repository: \url{https://github.com/aarunku5/Image-or-Information-Vis-2023} for visualization metadata, demographic details of study participants, study analysis and results.



\section*{Figure Credits}
\label{sec:figure_credits}
 A. Fig. \ref{fig:teaser} image credit: Amanda Shendruk / Quartz, January 2023
 
B. Fig. \ref{fig:3} image credit: WHO/UNICEF Joint Monitoring Programme for Water Supply and Sanitation / Progress on Drinking Water and Sanitation 2012-Update

C. Fig. \ref{fig:5} image credits (top-left to bottom-right for each score interval, separated by \textbf{;}): \textbf{I1:} MIT Senseable City Lab/ December 2010 \textbf{;} Teaching and Learning / Fortune Magazine, April 1961 \textbf{;} Paul Butler / Facebook, December 2010 \textbf{;} Sam Shannon / World Visualization Data Prize, Information is Beautiful February 2019 \textbf{;} Abhishek Tripathi / Mapbox, September 2019 \textbf{;} Steve Parker / GISP, Visual Wall Maps, January 2023. \textbf{I2:} Valentina D'Efilippo / Behance, 2014 \textbf{;} u/snickerdoodle\_codes / Reddit, r/dataisbeautiful, January 2023 \textbf{;} Audree Lapierre / Behance, December 2012 \textbf{;}  The Regional Planning Comission, County of Los Angeles / 1946\textbf{;}  Nathan Yau / Flowing Data, based on New York Times Hurricane Tracker, October 2012 \textbf{;} Moritz Stefaner / Truth \& Beauty in association with Google, Rhythm of Food, November 2016. \textbf{I3:} Nathan Yau / Flowing Data, August 2021 \textbf{;} Wm. F. Sheehan / University of Santa Clara, Ref. Chemistry, Vol. 49, No.3, p 17-18, 1976 \textbf{;} Kitchn Editors / Kitchn, October 2020 \textbf{;} Graphic Detail / Economist, October 2011  \textbf{;} Graphic Detail / Economist, August 2011 \textbf{;} David McCandless / Information is Beautiful, January 2019. \textbf{I4:} Nadieh Bremer / Visual Cinnamon for Scientific American, Graphic Science, July 2017 \textbf{;} Mike Bostock / Observable, October 2021 \textbf{;} Nadieh Bremer / Visual Cinnamon for The Guardian, December 2017 \textbf{;}  Beautiful News / Information is Beautiful, September 2020 \textbf{;} Eduardo Porter and Karl Russel  / New York Times, June 2018 \textbf{;}  Emily Badger, Claire Cain Miller, Adam Pearce, and Kevin Quealy / Upshot, New York Times, March 2018. \textbf{I5:} Jen Christiansen / Inflation Report, Bank of England, February 2010 \textbf{;} World Health Organization / Women's Health Report 2009 \textbf{;} Wall Street Journal / December 2011 \textbf{;} Hive Systems / March 2022 \textbf{;} Information is Beautiful / December 2019 \textbf{;} World Health Organization / Water and Sanitation, GLAAS Report, 2012.




\acknowledgments{
This research was supported in part by the U.S. National Science Foundation through grants DUE-2216452 and IIS-2238175.
}

\bibliographystyle{abbrv-doi-hyperref}

\bibliography{00_template}









\end{document}